\documentclass[aps,pra,nofootinbib,groupedaddress,twocolumn,final]{revtex4}

\usepackage{amssymb}
\usepackage{amsfonts}
\usepackage{amsmath}
\usepackage{verbatim}

\def\sb{{\sf b}}

\def\cH{{\cal H}}

\newcommand\ket[1]{| #1 \rangle}
\newcommand\bra[1]{\langle #1 |}
\newcommand\braket[2]{\langle #1|#2\rangle}
\newcommand\ketbra[2]{|#1\rangle\langle #2|}

\begin{document}

\title{Unconditionally Secure Quantum Bit Commitment\footnote{NOTE: This paper
is an expanded, self-contained version of QBC4 presented in quant-ph 0305143.
It is the basis of talks to be presented at the University of Pavia, Italy on
May 23, 2005 and Nottingham University, England on May 25, 2005.}}

\author{Horace P. Yuen}\email{yuen@ece.northwestern.edu}
\affiliation{Department of Electrical and Computer Engineering, Department of
Physics and Astronomy, Northwestern University, Evanston, IL
60208-3118, USA}




\begin{abstract}
The ``impossibility proof'' on unconditionally secure quantum bit
commitment is examined. It is shown that the possibility of 
juxtaposing quantum and classical randomness has not been properly
taken into account. A specific protocol that beats entanglement cheating
with entanglement is proved to be unconditionally secure.
\end{abstract}

\maketitle





Bit commitment is a kind of a cryptographic protocol that can serve as
a building block to achieve various cryptographic objectives, such as
user authentication. There is a nearly universal acceptance of the
general impossibility of secure quantum bit commitment (QBC), taken to
be a consequence of the Einstein-Podolsky-Rosen (EPR) type
entanglement cheating which supposedly rules out QBC and other quantum
protocols that have been proposed for various cryptographic objectives.
In a {\it bit commitment} scheme, one party, Adam,
provides another party, Babe, with a piece of evidence that he has
chosen a bit \sb\ (0 or 1) which is committed to her.  Later, Adam
would {\it open} the commitment by revealing the bit \sb\ to Babe and
convincing her that it is indeed the committed bit with the evidence
in her possession and whatever further evidence Adam then provides,
which she can {\it verify}.  The usual concrete example is for Adam to
write down the bit on a piece of paper, which is then locked in a safe
to be given to Babe, while keeping for himself the safe key that can
be presented later to open the commitment.  The scheme should be {\it
binding}, i.e., after Babe receives her evidence corresponding to a given bit
value, Adam should not be able to open a different one and convince
Babe to accept it. It should also be {\it concealing}, i.e., Babe
should not be able to tell from her evidence what the bit \sb\ is.
Otherwise, either Adam or Babe would be able to cheat successfully.

In standard cryptography, secure bit commitment is to be achieved
either through a trusted third party, or by invoking an unproved
assumption concerning the complexity of certain computational
problems.  By utilizing quantum effects, specifically the intrinsic
uncertainty of a quantum state, various QBC schemes not
involving a third party have been proposed to be unconditionally
secure (US), in the sense that neither Adam nor Babe could cheat with any
significant probability of success as a matter of physical laws.  In
1995-1996, a supposedly general proof of the impossibility of
unconditionally secure QBC, and the insecurity of previously proposed
protocols, were presented \cite{may1}-\cite{lc3}.  Henceforth it has
been generally accepted that secure QBC and related objectives are
impossible as a matter of principle \cite{lo}-\cite{sr}.

There is basically just one impossibility proof (IP), which gives the EPR
attacks for the cases of equal and unequal density operators that Babe
has for the two different bit values.  The proof purports to show that
if Babe's successful cheating probability $P^B_c$ is close to the
value 1/2, which is obtainable from pure guessing of the bit value,
then Adam's successful cheating probability $P^A_c$ is close to the
perfect value 1.   The impossibility proof
describes the EPR attack on a specific type of protocols, and then
argues that all possible QBC protocols are of this type. Since there is
no mathematical characterization of all possible QBC protocols - no mathematical
definition of a QBC protocol exists with the justification that it 
includes all protocols that would achieve bit commitment - a priori there
can be no general impossibility proof. A general analysis of the situation
is provided in \cite{yuen1}. In this paper, we pinpoint the gaps in the IP
involving quantum versus classical randomness that make possible a relatively
simple QBC protocol that utilizes classical random numbers generated in any
usual way. This particular protocol depends critically on verifying split
entangled pairs used as {\it anonymous states} \cite{yuen2,yuen3} which
Babe first transmitted to Adam in a two-stage protocol, thus beating
entanglement with entanglement \cite{yued}.

The impossibility proof, in its claimed generality, has never been
systematically spelled out in one place, but the essential ideas that
constitute this proof are generally agreed upon
\cite{may2}-\cite{sr}.  The formulation and the proof can be cast as
follows.  Adam and Babe have available to them two-way quantum
communications that terminate in a finite number of exchanges, during
which either party can perform any operation allowed by the laws of
quantum physics, all processes ideally accomplished with no
imperfection of any kind.  During these exchanges, Adam would have
committed a bit with associated evidence to Babe.  It is argued that,
at the end of the commitment phase, there is an entangled pure state
$\ket{\Phi_\sb}$, $\sb \in \{0,1\}$, shared between Adam who
possesses state space $\cH^A$, and Babe who possesses $\cH^B$.  For
example, if Adam sends Babe one of $M$ possible states $\{
\ket{\phi_{\sb i}} \}$ for bit \sb\ with probability $p_{\sb i}$, then
\begin{equation}
\ket{\Phi_{\sb }} = \sum_i \sqrt{p_{\sb i}}\ket{e_i}\ket{\phi_{\sb i}}
\label{eq:entstate}
\end{equation}
with orthonormal $\ket{e_i} \in \cH^A$ and known $\ket{\phi_{\sb i}}
\in \cH^B$.  Adam would open by making a measurement on $\cH^A$, say
$\{ \ket{e_i} \}$, communicating to Babe his result $i_0$ and $\sb$;
then Babe would verify by measuring
the corresponding projector $\ketbra{\phi_{\sb i_0}}{\phi_{\sb i_0}}$ on $\cH^B$,
accepting as correct only the result 1. More generally, one may
consider the whole $\ket{\Phi_\sb}$ of (\ref{eq:entstate}) as the
state corresponding to the bit $\sb$, with Adam sending $\cH^A$ to
Babe upon opening, so she can verify by projection measurement on $\ketbra{\Phi_\sb}{\Phi_\sb}$.

Classical random numbers are routinely used in classical cryptographic
protocols, and so must be allowed in a quantum protocol. In the IP, they
are handled as follows. When classical random numbers known only to one party
are used in the commitment, they are to be replaced by corresponding
quantum state purification.  The commitment of $\ket{\phi_{\sb
    i}}$ with probability $p_{\sb i}$ in (\ref{eq:entstate}) is, in
fact, an example of such purification. 
Generally, for any random $k$ used by Babe, it is argued that from the
doctrine of the ``Church of the Larger Hilbert Space'' \cite{gl}, it is to be
replaced by the purification $\ket{\Psi}$ in $\cH^B\otimes \cH^B$,
\begin{equation}
\ket{\Psi} = \sum_k \sqrt{\lambda_k} \ket{\psi_k}\ket{f_k},
\label{eq:purif}
\end{equation}
where $\ket{\psi_k} \in \cH^B$ and the $\ket{f_k}'s$ are complete orthonormal in $\cH^B$ kept
by Babe while $\cH^C$ would be sent to Adam.  With such purification, it is claimed that any protocol
involving classical secret parameters would become quantum-mechanically
determinate, i.e., the shared state $\ket{\Phi_\sb}$ at the end of
commitment is completely known to both parties.  This means that both $\{\lambda_k\}$ and
$\{\ket{f_k}\}$ are taken to be known exactly to both Babe {\it and
Adam}. The IP assumes that Babe is honest (and Adam is also honest in
a multi-stage protocol \cite{sr,comm1},\cite{lc1}) in using the 
agreed upon $\{\lambda_k\}$ and $\{\ket{f_k}\}$, and then claims
that unconditional security is impossible. We will retain this
assumption in this paper to show that the IP reasoning is incorrect.
However, US QBC is possible even when this assumption is dropped
by using a cheat-testing procedure \cite{yuen3,yued}.

In the purification (2), exactly which orthonormal $\{\ket{f_k}\}$ 
is used does not affect the anonymous nature of $\{\ket{\psi_k}\}$.
Why then does $\{\ket{f_k}\}$ have to be agreed upon and known to
Adam? This issue is {\it not} addressed in the IP. Clearly, a choice
from a set of possible $\{\ket{f_k^l}\},l \in \{1,\cdots,L\}$ with
a priori probabilities $\{p^l\}$, both openly known, can be picked
secretly by Babe using a classical random number generator for each
transmission of $\ket{\Psi}$. If the protcol is concealing
for every $l$, Adam has no right to demand the knowledge of $l$. On
the other hand, the IP may {\it not} go through as $\{\ket{f_k}\}$ or the
total 
\begin{equation}
\ket{\Phi_{\sb}} = \sum_{ik}\sqrt{p_{\sb i}\lambda_k} \ket{e_i}
\ket{f_k} \ket{\phi_{\sb ik}}
\label{eq:totphi}
\end{equation}
is {\it not} known to Adam. This is the anonymous state idea \cite{yuen2,yuen3}
for building US QBC protocols. It should be noted that if Babe, e.g., picks
$\{\ket{\psi_k}\}$ by throwing a die with probabilities $\{\lambda_k\}$, she
herself would not be able to tell what the $\{\ket{f_k}\}$ is. Thus, physically,
it is totally unreasonable to assume that Adam knows the $\{\ket{f_k}\}$
in an anonymous state protocol. Similarly, as just noted, Babe may just send
a classically randomly chosen $\{\ket{\psi_k}\}$ so long as the protocol 
is concealing for every $k$. As it turns out \cite{yuen3}, if the protocol is
perfectly concealing ($P_c^B = 1/2$),  Adam's cheating
transformation $U^A$ on $\cH^{A}$ that brings $\ket{\Phi_0}$ to
$\ket{\Phi_1}= U^A \otimes I^B \ket{\Phi_0}$ is independent of $\{\ket{f_k}\}$
or the specific $\ket{\psi_k}$, under either of the following conditions:
(a) Babe verifies by first measuring
$\{\ket{f_k}\}$ and then checking $\ket{\psi_k}$, or  (b) Adam's \sb-dependent
commitment action does not change the composite index $k$ to get one unknown
state to another unknown state for him. One way, among others, 
to show this is
to use the result in \cite{yuen3} that explicitly determines $U^A$ 
in terms of $\ket{\Phi_{\sb}}$ of (1) or (3), which can be achieved
by a simple matrix transformation argument \cite{note}. Let $U_{ij} \equiv
\bra{e_i}U^A\ket{e_j}, \Lambda_{ji}\equiv\sqrt{p_{i}p_{j}}\braket{\phi_{1i}}{\phi_{0j}},
|\Lambda|=(\Lambda\Lambda^{\dag})^{1/2}.$ Then \cite{yuen3,note}
\begin{equation}
\Lambda U=|\Lambda|.
\label{eq:polar}
\end{equation}
Generalization to $\epsilon$-concealing ($P_c^B = 1/2+\epsilon $) protocols
of this behavior can be expected.

However, a perfectly concealing US protocol may be obtained from the use
of (2) with $\ket{\psi_k}$ being an entangled state split between Adam
and Babe during commitment with Adam's \sb-dependent commitment action
changing the composite index $k$, while verification is carried out on the 
total entangled $\ket{\psi_k}$. Such a split entangled state by
itself does not lead to a binding protocol for known$\{\ket{f_k}\}$,
but together with the use of a secretly chosen $\ket{f_k^l}$ as 
described above, Adam would not be able to cheat perfectly ($P_c^A = 1$).
Thus, an $\epsilon$-binding protocol for any $\epsilon>0$ is obtained in
a sufficiently long $n$-sequence in the standard fashion \cite{lc1}. In the
following, we describe the specific protocol (which we call QBC4) that
achieves unconditional security in the above fashion.

Let $\ket{m_j}_j$, $j \in \{ \mu,\nu\}$, $m_j \in \{1,2\}$, be two openly 
known orthonormal qubit
states, $\braket{1}{2}=0$, for each of the two possible $j$. When
there is no ambiguity, we would write $\ket{m_j}_j$ simply as
$\ket{m}_j$ to simplify notation. Let Babe prepare two states
\begin{equation}
\ket{\Psi_j} = \frac{1}{\sqrt{2}} \sum_m \ket{m}_j \ket{g_m}_j,
\label{eq:psij}
\end{equation}
where $\ket{m}_j \in \cH^B_{j\alpha}$, $m \in \{1,2\}$, and
$\{\ket{g_m}_j | m=1,2\}$ form an orthonormal basis in
$\cH^B_{j\beta}$ for each $j \in \{\mu,\nu\}$, with $\ket{\Psi_j} \in
\cH^B_{j\alpha} \otimes \cH^B_{j\beta}$ on two qubits for each
$j$. We have skipped one subscript $j$ in $\ket{g_{m_j}}_j$ as in
$\ket{m}_j$ to simplify notation. Let $\cH^B_\alpha \equiv
\cH^B_{\mu\alpha} \otimes \cH^B_{\nu\alpha}$, $\cH^B_\beta \equiv
\cH^B_{\mu\beta} \otimes \cH^B_{\nu\beta}$, $\cH^B \equiv \cH^B_\alpha
\otimes \cH^B_\beta$.

Babe keeps $\cH^B_\beta$ and sends the ordered pair of qubits
$\cH^B_\alpha$ to Adam. Adam applies the following transformation on
$\cH^B_{j\alpha}$ separately for each $j$: $\ket{\Psi_j}$ becomes
$\ket{\Phi_j} \in \cH^A_j \otimes \cH^B_{j\alpha} \otimes
\cH^B_{j\beta}$:
\begin{equation}
\ket{\Phi_j} = \frac{1}{\sqrt{8}} \sum_{m,i} \ket{e_i}_j V_i \ket{m}_j
\ket{g_m}_j,
\label{eq:phij}
\end{equation}
where $i \in \{1,2,3,4\}$, $\{\ket{e_i}_j\}$ complete orthonormal in
$\cH^A_j$, and $V_i$ are four unitary qubit operators given by $I$,
$\sigma_x$, $-i\sigma_y$, $\sigma_z$ in terms of the Pauli spin
operators when $\ket{1}$ and $\ket{2}$ lie on the qubit
$z$-axis. Eq.~(\ref{eq:phij}) can be obtained by the unitary
transformation $\sum_i \ket{e_i}_{jj}\bra{e_i} \otimes V_i$ on $\cH^A
\otimes \cH^B_{j\alpha}$ with initial state $\ket{\psi_A} \in \cH^A$
that has $\braket{e_i}{\psi_A}_j = \frac{1}{2}$. To commit $\sb=0$, Adam
sends back $\cH^B_{\mu\alpha} \otimes \cH^B_{\nu\alpha}$ in the
original order, and he switches them to $\cH^B_{\nu\alpha}\otimes
\cH^B_{\mu\alpha}$ to commit $\sb=1$. He opens by announcing $\sb$,
the order of the two $\cH^B_{j\alpha}$ he committed, and submitting
the ordered qubit pair $\cH^A \equiv \cH^A_\mu \otimes
\cH^A_\nu$. Babe verifies by measuring the corresponding projections
to $\ket{\Phi_\mu}\ket{\Phi_\nu}$ of (\ref{eq:phij}). The general
situation is depicted in Fig. 1.

It is easy to verify by tracing over $\cH^A$ that for either $\sb$,
$\rho^B_0 = \rho^B_1 = I^B/16$ on $\cH^B$, for any orthonormal
$\{\ket{g_m}_j\}$. If Babe entangles over the possible choices of such
$\{ \ket{g_m}_j\}$ via $\{\ket{f_k}\}$, a simple calculation shows that perfect concealing
$\rho^{BC}_0 = \rho^{BC}_1$ on $\cH^B \otimes \cH^C$ is maintained,
where $\cH^C$ is the space Babe used to carry out such
entanglement. Similarly, pefect concealing is maintained with further
entanglement of $\{\ket{f_k^l}\}$ with $\{p^l\}$. This happens because the $V_i$ operations by Adam
totally disentangle the state on $\cH^B_\alpha \otimes \cH^B_\beta
\otimes \cH^C$ into a product state $I^B_\alpha / 4 \otimes
\rho^{BC}_\beta$ for either $\sb$, and there is no identity that
individuates a qubit by itself, that is not entangled or correlated to another. 

Intuitively, we intend to guarantee binding by the fact that
$\cH^B_{j\beta} = \cH^B_{\mu\beta} \otimes \cH^B_{\nu\beta}$ in Babe's
possession cannot be switched to $\cH^B_{\nu\beta}\otimes
\cH^B_{\mu\beta}$ by operating on $\cH^A \otimes \cH^B_\alpha$
alone. However, this is possible if the two orthonormal sets $\{
\ket{g_m}_j\}$ are known. Indeed, this is the content of the
impossibility proof \cite{comm2}. Thus, to guarantee security, Babe
needs to employ different choices of $\{\ket{g^{k'}_m}_j\}$ with
different bases indexed by $k'$. She may employ a fixed probability
distribution $\{ p_{k'j}\}$ for each $j$, and entangle these
via orthonormal $\{ \ket{g^{k'}}_j\}$, ad infinitum. This possible chain
of purifications has to stop somewhere, and we simply stop it at
$\cH^B$ without $\cH^C$. As we have seen, this does not affect perfect
concealing so that Babe is free to choose any orthonormal
$\{\ket{g_m}_j\}$. In the notation of (2), the effective $\{\ket{\psi_k}\}$
in this case is $\ket{\Psi_{\mu}^{k'}}\ket{\Psi_{\nu}^{k'}}$ determined
by $\{\ket{g_m^{k'}}\}$, with further entanglement to $\ket{f_k}$ of (2)
described by $\cH^{C}$ above. In the notation of (2), $k$ is the ordered
triple $(\mu,\nu, k')$ for fixed $\{\ket{g_m^{k'}}\}$. 
It is clearly unreasonable for Adam to demand such
knowledge, as discussed above and codified in the Secrecy Principle of 
Ref.~\cite{comm2}. This possibility
is neglected in the impossibility proof.

To see exactly how binding is obtained in the present situation, note
that the perfect cheating transformation $U^A$ is determined by
Eq.~(4), which is unique up to a phase factor in
this nondegenerate situation. It depends on 
$\{\ket{g_m}_j\}$ in the present case with
state-space switching, i.e. $\mu,\nu$-switching where $\{\mu,\nu\}$ is part of the composite
index $k$, in contrast to merely $\braket{g_m}{g_{m'}} =
\delta_{mm'}$, i.e., no dependence on the actual $\{\ket{g_m}\}$ in
the case without switching in the absence of $j$. Thus, Adam
cannot cheat perfectly. Note that the generalized IP result from \cite{yuen3}
does not apply here because Adam's \sb-dependent commitment action re-arranges
the $\mu,\nu$ part of the composite index $k$ of (2), which in turn demands
entanglement or correlation from Babe in order that she can verify such re-arrangement.
On the other hand, quantum entanglement instead of classical correlation is also
needed here -- Babe cannot verify by first measuring $\{\ket{m}_j\}$ because Adam
would be able to determine the $\ket{m}_j$ with a measurement if he knows
that is the way Babe would verify. Thus, we are indeed beating
entanglement with entanglement.  
On the other hand, Adam's entanglement is not
essential. As usual in QBC protocols, the whole procedure works the
same if Adam chooses the $V_i$ on $\cH^B_{\mu\alpha}$ and
$\cH^B_{\nu\alpha}$ classically and opens by telling Babe his choice.

We have assumed as usual that Adam opens $\sb=0$ perfectly. Let $p_A <
1$ be Adam's optimum probability of cheating for a given choice of 
$\{ \ket{g^{k'}_m}_j\}$ and $\{
p_{k'j}\}$, taking into account also all his other obvious imperfect
cheating possibilities, such as simply announcing a different
$\sb$. We have thus shown that the formulation and the reasoning of
the impossibility proof break down already in this simple pair
$\ket{\Phi_\mu}\ket{\Phi_\nu}$ situation.

When $\sb=0$ pefect opening condition is relaxed, it is clear that
Adam still cannot cheat perfectly, but it is possible that the overall
successful opening probability (honest plus cheating) may be improved. By continuity it can be seen that Adam's optimum
cheating probability $\bar{P}^A_c$ is arbitrarily close to $p_A=\frac{1}{2}$ if
the $\sb=0$ opening probability is arbitrarily close to $1$, the case
of interest.

Protocol QBC4 is obtained when the above protocol, to be called QBC4p, 
is extended to a sequence of $\{\ket{\Psi_{n \mu}}\ket{\Psi_{n \nu}}\}$, 
$n \in \{1,\ldots,N\}$,
each of the form (\ref{eq:psij}), with $\ket{g_{nm}}_j \in
\cH^B_{nj \beta}$, $\ket{m_n}_j \in \cH^B_{nj \beta}$,
etc. Babe should send Adam $\{ \cH^B_{n\mu \alpha} \otimes \cH^B_{n
\nu \alpha}\}$ and Adam should commit to Babe these spaces for all
$\mu$ after he entangles them with $\cH^A_{n\mu} \otimes
\cH^A_{n\nu}$ using the $V_i$ operations, permuting each pair for
$\sb=1$. He opens by announcing $\sb$ and the state of the qubits in
each $\cH^B_{n\alpha}$ and
submitting $\{\cH^A_n\}$, with Babe verifyng
$\ket{\Phi_{n\mu}}\ket{\Phi_{n\nu}} \in \cH^A_n \otimes
\cH^B_n$ after possible rearrangement for each $n$. Since there
is no new entanglement possibility for Adam, the protocol is perfectly
concealing with $\bar{P}^A_c = p^n_A$ going to zero exponentially in
$N$. Thus, QBC4 is perfectly concealing and $\epsilon$-binding for any
$\epsilon > 0$ by letting $N$ be large. We summarize our perfectly concealing and
$\epsilon$-binding protocol:

\begin{center}
\vskip 0.12in
\framebox{
\begin{minipage}{0.9\columnwidth}
\vskip 0.1in
\underline{PROTOCOL {\bf QBC4}}

{\small \begin{enumerate}
\item[(i)] Babe sends Adam $N$ ordered pairs $\{ \cH^B_{n \mu\alpha}
\otimes \cH^B_{n \nu\alpha} \}$ of qubit pairs, $n \in \{1,\ldots,N\}$,
which are entangled to $\{ \cH^B_{n \mu\beta} \otimes \cH^B_{n \nu\beta}\}$
in her possession in states $\ket{\Psi_{n \mu}}\ket{\Psi_{n \nu}}$
of the form (\ref{eq:psij}), with independent random choices of
$\{\ket{g^{k'}_m}_j\}$ with probability $\{p_{k'j}\}$.
\item[(ii)] To commit $\sb$, Adam applies, for each $n$, $\sum_i
\ketbra{e_i}{e_i} \otimes V_i$ on $\cH^A_n \otimes \cH^B_{n
  \alpha}$, resulting in a state
$\ket{\Phi_{n\mu}}\ket{\Phi_{n\nu}}$ given via
the form (\ref{eq:phij}), and sends $\{ \cH^B_{n \alpha}\}$ to
Babe as evidence for $\sb=0$, while switching the order ot each
$\cH^B_{n\mu\alpha}\otimes \cH^B_{n\nu\alpha}$ for $\sb=1$.
\item[(iii)] Adam opens by announcing $\sb$, the order of the qubits
  in each $\cH^B_{n\alpha}$, and submitting $\{
\cH^A_n\}$. Babe verifies by projective measurements of $\{
\ket{\Phi_{n\mu}}\}$, $\{\ket{\Phi_{n\nu}}\}$, for all $n$.
\end{enumerate}
\vskip 0.1in
}
\end{minipage}
}
\end{center}
\vskip 0.15in


In conclusion, the possibility of unconditionally secure quantum bit commitment
opens up the possibility of many cryptographic functions, including secure 
multi-party computation. It would be of interest to develop practically feasible
secure QBC protocols \cite{yuen1} for such applications.
\begin{acknowledgments}

I would like to thank G.M.~D'Ariano, W.Y.~Hwang, and R.~Nair for useful discussions. 
This work was supported by the Defense Advanced Research Projects
Agency and by the U.S. Army Research Office.

\end{acknowledgments}


\begin{thebibliography}{99}
\bibitem{note}The maximum of $|\bra{\Phi_1} U^A \ket{\Phi_0}|=|tr U\Lambda|$
over all unitary $U$ is obtained when $U\Lambda$ is nonnegative definite with
maximum value given by $tr |\Lambda|=\bar{P}^A_c$. Thus U is determined by the
polar decomposition of $\Lambda=|\Lambda|U^{\dag}$.
\end{thebibliography}
\end{document}